# On nonlinear graphene response on monochromatic electromagnetic wave in the form as generation of harmonics

Michael V. Davidovich

Saratov National Research State University named after N. G. Chernyshevsky,
Russia, 410012, Saratov, Astrakhan str., 83
E-mail: davidovichmv@info.sgu.ru

**Abstract.** We consider the linear and nonlinear response of a weighted graphene sheet under the normal incidence of a plane electromagnetic wave in the form of a quasi-monochromatic pulse of long duration with a sharp edge and harmonic filling. The generation of odd harmonics in the reflected and transmitted spectra is obtained. The coefficient of transformation of the first harmonic into the third harmonic at a frequency of 10 Hz is of the order of $10^{-3}$. We use perturbative theory based on the quantum Wallace strong coupling model with field amplitude expansion and integration over the entire Brillouin zone (BZ), solving the kinetic Boltzmann equation with a collision integral in the Bhatnagar-Gross-Crook (BGK) form. The electromagnetic field is considered classically, while its vector potential changes the quasi-pulse in the dispersion equation.

## 1. Introduction

The band structure of graphene was first obtained as a solution of the Schrodinger equation (SE) with a periodic potential in the strong coupling approximation in [1], and generally has the form of two overlapping $\pi^{\pm}$ subbands in the form of [2-4] $E^{\pm}(\mathbf{q}) = (\varepsilon_{2p} \pm \gamma_0 w(\mathbf{q}))/(1 \mp \gamma_1 w) \approx \pm \gamma_0 w(\mathbf{q}) + \gamma_1 \gamma_0 w^2(\mathbf{q}) + \varepsilon_{2p}$. Here, the coupling constant (nearest neighbor energy overlap) $2.5 \leq \gamma_0 \leq 3$ (eV) is significantly greater than the overlap integral $\gamma_1 < 0.1$ and $\varepsilon_{2p} \approx 0$ eV. They are usually determined from experiment, or from the first principles of calculations (usually this is the method of density functional theory). Assuming $\gamma_1 = 0$, we obtain the commonly used symmetric BZ structure [5] (see also [1-14]):

$$w(\mathbf{q}) = \sqrt{1 + 4\cos(q_x a\sqrt{3}/2)\cos(q_y a/2) + 4\cos^2(q_y a/2)}. \qquad (1)$$

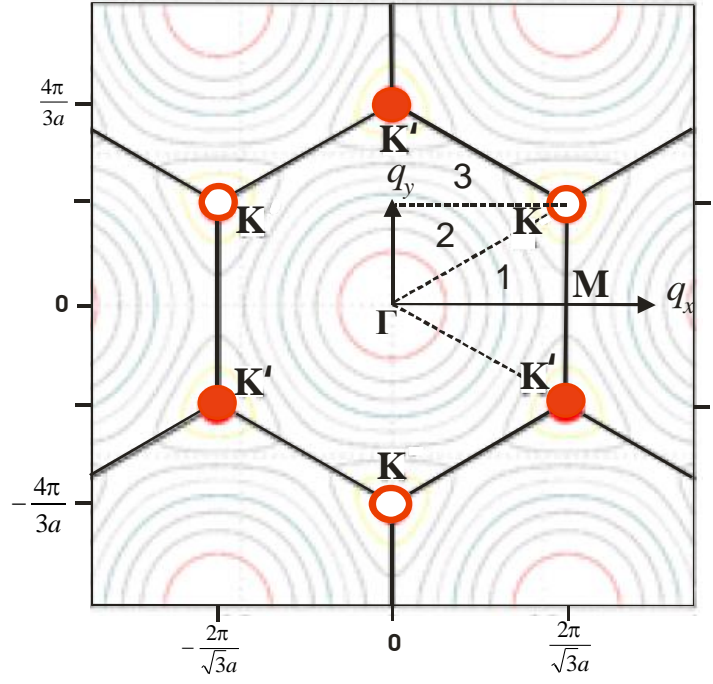

Fig. 1. Two-dimensional band structure of graphene with equipotential level lines

Here $\mathbf{q} = \mathbf{p}/\hbar$, $\mathbf{p}$ is the 2D quasi-pulse, therefore, we also denote $w(\mathbf{p}) = w(p_x, p_y)$. BZ (1) gets the opportunity to control the 2D Hamiltonian

$$H(\mathbf{q}) = \begin{pmatrix} 0 & h(\mathbf{q}) \\ h^*(\mathbf{q}) & 0 \end{pmatrix}, \qquad (2)$$

where $h(\mathbf{q}) = \gamma_0 \left[ \exp(iq_x a) + 2\exp(-iq_x a/2)\cos(q_y a\sqrt{3}/2) \right]$, $a = 0.246$ nm is the lattice constant of graphene. We take $\gamma_0 = 3$ eV, i.e. the Fermi velocity is $v_F = \sqrt{3}\gamma_0 a/(2\hbar) = 0.9739 \cdot 10^6$ m/s. BZ (1) is shown in Fig. 1 [1]. Graphene is a highly nonlinear material. Its nonlinear properties have been studied theoretically [15-38] and experimentally [37-51] in a large number of papers. The generation of odd harmonics was detected [43-46]. Most theoretical studies are based on the use of the quantum electrodynamics (QED) model of massless Dirac fermions in the approximation of linear dispersion $E(\tilde{\mathbf{q}}) = \pm \hbar |\tilde{\mathbf{q}}| v_F$ with valley $g_v = 2$ and spin $g_s = 2$ degeneracy. In this model, the wavenumber $\tilde{\mathbf{q}} = \mathbf{q} - \mathbf{q_D}$ and energy are calculated from the zero Fermi level (FL) corresponding to two Dirac points $\mathbf{K}$ and $\mathbf{K}'$ (Fig.1). The formal similarity with QED when a strong field is applied to a quantum vacuum has led to a large number of works on nonperturbative methods for determining the generation of quasi-particle pairs such as the Schwinger effect [23], as well as for Landau-Zener tunneling [32].

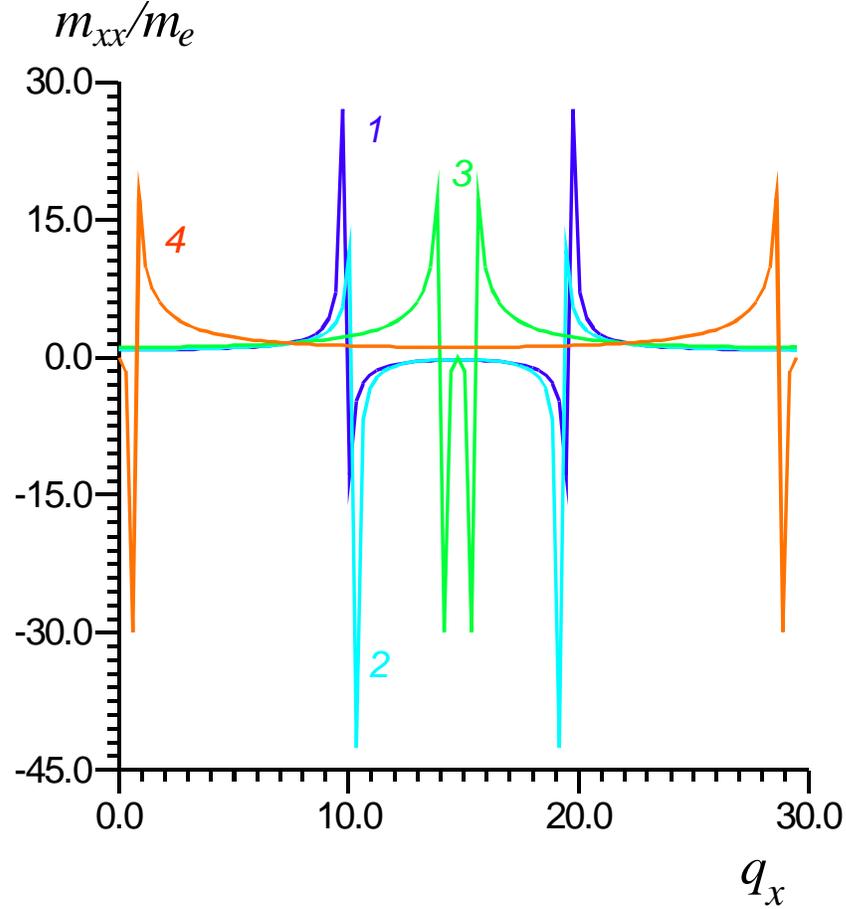

Fig. 2. Dependence of the normalized effective mass $m_{xx}(\mathbf{q})/m_e$ on $q_x$ at different $q_y$: $q_y = 0$ (curve 1), $q_y = \pi/(3a)$ (2), $q_y = 2\pi/(3a)$ (3), $q_y = 4\pi/(3a)$ (4). The peaks correspond to the poles. Near the Dirac points, the tendency to zero is almost linear (curve 3, $q_x \approx 2\pi/(\sqrt{3}a)$, curve 4 at $q_x \approx 0$ and $q_x \approx 4\pi/(\sqrt{3}a)$ )

It should be noted that graphene differs significantly from the quantum vacuum model. At the Dirac points, the density of states (DOS) $D$ and the effective and mass both are zero, and when displaced from them, a nonzero effective mass $\hat{m}(\mathbf{q}) = \hbar^2 (\nabla_{\mathbf{q}} w(\mathbf{q}))^{-1}$ appears (see Fig. 2), which calls into question the accuracy of models of massless fermions. As the Dirac points approach, the components of the mass tensor tend to zero, but only in their very small neighborhood. In the BZ, they also have poles. The particles can be located not only in valleys (Dirac cones), but also in the vicinity of the BZ center, therefore, estimating their number from the DOS and in the massless model gives a value $D(E) = g_s g_v |E|/(2\pi\hbar^2 v_F^2)$ [52] 1.6 times greater than $n_0 = 3.8 \cdot 10^{19}$ 1/m$^2$ (corresponding to two π electrons per graphene hexagon with the area $S_0 = \sqrt{3}a^2/2 = 0.05238$ nm$^2$). This is due to the linear approximation of $D$, which is overestimated in comparison, for example, with the calculation from the first principles of calculations. Thus, valley degeneration $g_v = 2$ leads to an overestimation of DOS. QED uses electrical neutrality, which for the distribution functions (DF) in graphene is interpreted as the electrical neutrality of the electronic and hole subsystems $f(\mathbf{p},t) = f_e(\mathbf{p},t) = f_h(\mathbf{p},t)$ [36]. In fact, this is not the case: $f(\mathbf{p},t) = f_e(\mathbf{p},t) = 1 - f_h(\mathbf{p},t)$ [1,5], which corresponds to the conservation of the number of particles (the main condition for normalization in quantum mechanics). At zero temperature, the conduction band (CB) $\pi^+$ is empty [1], all electrons are in the valence band (VB), and electrical neutrality is provided by carbon ions in the lattice. In these papers, strong unsteady fields and infinite Dirac cones are considered. The BZ based on (1) was obtained for a stationary case at zero temperature and in the absence of disturbances. A time-dependent strong field changes the BZ of Fig. 3, leads to heating of the carriers, and with their energy in the field of the order of $\gamma_0$, to the breaking of π bonds with the transformation of graphene into a plasma layer. Dirac cones have a size of about 1 eV, and then smoothly transition from the CB band to the high-energy part of the $\pi^+$ zone. The same applies to the $\pi^-$ zone. In models, they are considered infinite, i.e. at energies of the order of 1 eV or more, the linear dispersion gives a significant error. This limits the range of applicability of such models to the THZ range. All processes with a high field energy density require non-stationary consideration. Next, let's look at the conditions under which the BZ is strongly distorted. In the strongly nonstationary case, a nonstationary SE with a periodic graphene potential should be solved. The energy is not defined for such case. Since graphene is in thermodynamic equilibrium with the thermal field, its wave function has not been determined, therefore, the Neumann equations for the density matrix should be solved and the current density should be determined as a function of time [53]. In addition, if the energy of quanta $\omega\hbar \geq \gamma_0$ (which is also possible for short high-power pulses with a significantly lower carrier frequency), then the breaking of π-coupling also occurs, and a plasma layer model should be used to simulate the high-frequency response.

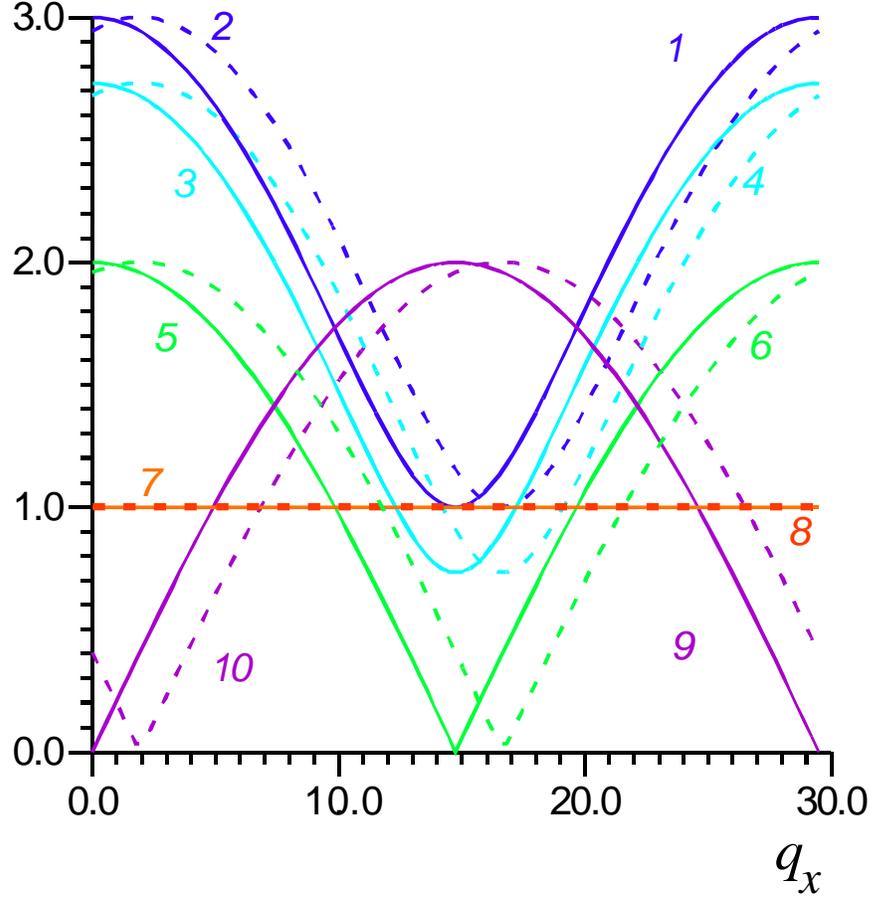

Fig. 3. Dependence of $w(\mathbf{q})$ (solid lines) and $w(t,\mathbf{q})$ (dashed lines) on $q_x$ (1/nm) at $q_0=1.91$ 1/nm ($A_x=1$A) and different $q_y$: $q_y = 0$ (curves 1,2), $q_y = \pi/(3a)$ (3,4), $q_y = 2\pi/(3a)$ (5,6), $q_y = \pi/a$ (7,8), $q_y = 4\pi/(3a)$ (9,10)

Nevertheless, for modeling kinetic processes in graphene, BZ in the form of Dirac cones is often used, for example, [15-32]. In fact, these cones do not have round, but triangular (with smoothed corners) cross-sections, Fig. 1. Moreover, for energies greater than 1 eV, they transform into a curved surface of the center of the BZ, i.e. for the optical range, the linear dispersion gives a significant error. Only in a very small neighborhood of Dirac points are equipotentials close to circles. The surface (1) and its complex relief are given in many works, for example, [51]. In [33], the nonperturbative kinetic equation QED [28-30] was applied to type (1) dispersion with quasi-pulse substitution $\mathbf{p} \to \mathbf{p} - \eta_0 e\mathbf{A}/c$. The impedance $\eta_0 = \sqrt{\mu_0/\varepsilon_0}$ is indicated here and below. In [24], such a substitution was used for the nonstationary Dirac equation with a time-linear vector potential $\mathbf{A}$, i.e. with a time-constant electric field. In [34], a general quantum theory of nonlinear local dynamic third-order conductivity for linear dispersion is developed. In [35], phenomenological relaxation parameters for third-order optical nonlinearity in doped graphene were investigated using a perturbative solution of the semiconductor Bloch equation at Dirac points. The static fields at which bonding phenomena occur in graphene are very high: about $10^{10}$ V/m. However, with fields 2-3 orders of magnitude lower, the BZ is significantly distorted, Fig. 3. The nonlinear properties of graphene are already manifested in rather weak fields, in which the heating of the carriers and the change in the chemical potential can be neglected. In harmonic fields, charge carriers oscillate and accelerate only for half a period, i.e. their range decreases inversely proportional to the square of the frequency with increasing frequency. At the same time $\mathbf{E}(\omega) \sim \omega \mathbf{A}(\omega)$. The use of linear dispersion $E(\tilde{\mathbf{q}}) = \pm |\tilde{\mathbf{p}}| v_F$ instead of dispersion $E(\mathbf{q}) = \pm \gamma_0 w(\mathbf{q})$ in the above studies implies that quasiparticles are located only in small neighborhoods of Dirac cones (two valleys). Although large energies are needed to locate particles in the vicinity of the BZ center, where their effective masses $m_{xx}^{\pm}(0) = m_{yy}^{\pm}(0) = \mp 0.278 m_e$ are finite and equal, it is important to take such processes into account. Since even with a small deviation from the Dirac points, the effective masses are finite, this suggests that it is advisable to build models using the entire BZ. These arguments suggest the expediency of constructing models of linear and nonlinear response of graphene based on the BZ method of tight-binding (1). In this paper, we use such a model. We believe that the field is not strong and weakly changes the band structure (1), and also does not lead to heating of carriers and to a change in the chemical potential $\mu_c$. In addition, we believe that the field acts for a long time and is quasi-monochromatic, i.e. Spectral values and a band structure can be used.

**2. Problem statement**

Linear response of an infinite graphene sheet lying in the $z=0$ plane to a monochromatic wave determined by the vector potential $\mathbf{A}(t,\mathbf{r}) = -\mathbf{A}_0 \Pi(t - z/c) \sin(\omega_0 t - \mathbf{kr}/c)$ and the electric field

$$\mathbf{E}(t,\mathbf{r}) = \mathbf{E}_0 \cos(\omega_0 t - \mathbf{kr}/c) \tag{3}$$

is given as the surface current density (hereinafter the word density will be omitted) in the form $j_\alpha(\omega, x, y) = \sigma_{\alpha\beta}(\omega, \mathbf{k}) E_\beta(\omega, x, y, 0)$, $\alpha, \beta = x, y$, where [6]

$$\sigma_{\alpha\beta}(\omega,\mathbf{k}) = \frac{ie^2}{\pi^2\hbar}\left\{\sum_{n=1,2}\int_{BZ}\frac{d^2p\,v_\alpha v_\beta\{f_0(\varepsilon_n(\mathbf{p}^-))-f_0(\varepsilon_n(\mathbf{p}^+))\}}{[\varepsilon_n(\mathbf{p}^+)-\varepsilon_n(\mathbf{p}^-)](\omega\hbar-\varepsilon_n(\mathbf{p}^+)+\varepsilon_n(\mathbf{p}^-))} + \right.$$
$$\left. + 2\omega\hbar\int_{BZ}\frac{d^2p\,v_\alpha^{12}v_\beta^{21}\{f_0(\varepsilon_1(\mathbf{p}^-))-f_0(\varepsilon_2(\mathbf{p}^+))\}}{[\varepsilon_2(\mathbf{p}^+)-\varepsilon_1(\mathbf{p}^-)](\omega^2\hbar^2-[\varepsilon_2(\mathbf{p}^+)-\varepsilon_1(\mathbf{p}^-)]^2)}\right\} \quad (4)$$

In (4) $\varepsilon_n(\mathbf{p}) = \varepsilon_n(\mathbf{q}) = (-1)^n \gamma_0 w(\mathbf{q})$, $n=1$ corresponds to the VB $\pi^-$ of the BZ, $n=2$ corresponds to the CB $\pi^+$, the velocity components $v_\alpha^\pm(\mathbf{q}) = \pm\gamma_0(\partial/\partial q_\alpha)w(\mathbf{q})/\hbar$, $\mathbf{p}^\pm = \mathbf{p} \pm \hbar\mathbf{k}/2 = \hbar(\mathbf{q} \pm \mathbf{k}/2)$, i.e. the spatial dispersion is taken into account. The first term in tensor (4) corresponds to the intraband conductivity, and the second term corresponds to the interband conductivity. Tensor (4) acts on the spectral amplitude of the electric field $E_\beta(\omega,x,y,0)$ on graphene. We consider the case $\mathbf{A}_0 = \mathbf{x}_0 A_0$ without limitation of generality and take the form-factor $\Pi(z) = \chi(z)$ as the Heaviside function. Then the quasi-monochromatic case corresponds $t \gg 2\pi/\omega_0$, i.e. the case of large times. Accordingly, at the moment $t=0$, the wave (3) with amplitude $\mathbf{E}_0 = \mathbf{x}_0 \eta_0 A_0/c$ approached graphene, and we consider fields and currents at long times. For such a wave with the amplitude $E_0 = \eta_0 A_0/c$ we have $E_{0x} = E_0 k_z/\sqrt{k_x^2+k_z^2}$, $E_{0z} = E_0 k_x/\sqrt{k_x^2+k_z^2} = E_0 k_x/k_0$. Let 's denote $\tilde{k} = k_x/k_0 = \sqrt{1-(k_x/k_0)^2}$. Because $k_y = 0$, the spatial dispersion is determined only by the component $k_x$, and with monochromatic exposure in a weak field, there is the linear response $j_x(\omega,x,y) = \sigma_{xx}(\omega,k_x)E_x(\omega,k_x x,y,0) = \sigma_{xx}(\omega,k_x)E_0\tilde{k}$, where the full field is indicated. During the propagation of plasmons, SD is determined by all components. A complex amplitude $\dot{\mathbf{E}}(t,\mathbf{r}) = \mathbf{E}_0(\omega_0\mathbf{k})\exp(-i(\omega_0 t - \mathbf{k}\mathbf{r}/c))$ corresponds to the wave (3). With a long action time $t$, all instantaneous spectra are close to the spectra of a monochromatic wave, i.e. they have small deviations from it. Then the real part $\mathrm{Re}(\dot{\mathbf{E}}(t,\mathbf{r}))$ gives the field (3), i.e. there is no sine component. If a wave falls at an angle, its vector potential $\mathbf{A} = \mathbf{x}_0 A_x + \mathbf{z}_0 A_z$ acts on graphene, but only its component $A_x(t) = -cE_0 \sin(\omega_0 t)\sqrt{1-(k_x/k_0)^2}/(\eta_0\omega_0)$. It changes the value of the quasi-pulse $p_x$ by $p_x - e\eta_0 A_x(t)/c$, which is reflected in (1) and (4). Therefore, it is sufficient to consider the normal drop when the SD does not need to be taken into account. It should be taken into account when spreading slow plasmons when $|\mathbf{k}| \gg k_0$. We enter the parameter $q_0(t) = e\eta_0 A_x(t)/(c\hbar)$. Under its influence, the BZ changes and becomes a function of time: $w(t,\mathbf{q}) = w(q_x - q_0(t), q_y)$, where $w(q_x, q_y)$ means the function (1), Fig. 3. Accordingly, the spectral tensor (4) becomes a function of the vector potential component. In the first approximation $A_x(t) = -cE_0 \sin(\omega_0 t)\tilde{k}/(\eta_0 \omega_0)$ (the incident wave approximation), i.e. the spectral function (4) should be interpreted as an instantaneous time-dependent spectrum: $\sigma_{\alpha\beta}(t, \omega, k_x)$. However, the current creates a vector potential, which should be taken into account in the full potential, i.e. there is a self-acting current. Considering this requires iterative approaches to solving nonlinear equations. Graphene is in thermodynamic equilibrium with a large system – the thermal Planck field, i.e. it is not a closed system. Under the influence of the potential, the SE becomes nonstationary with undefined energy. Its wave function must be replaced by density matrix, which should be found. To a first approximation, its diagonal elements contain perturbed zone distribution functions $f^{\pm}(t)$ of $\pi^{\pm}$ zones. Formula (4) is obtained as a linear response in the Kubo approximation using Green's temperature functions and Matsubara's technique by averaging over the Gibbs ensemble. Moreover, only the proportional **A** term in the expression for the current operator $j_\alpha = -e^2 \eta_0 \langle \tilde{\psi}^+ | m_{\alpha\beta}^{-1} | \tilde{\psi} \rangle A_\beta / c$ in the interaction representation is taken into account. The nonlinear current density should be taken in the form [6,14] $j_\alpha = j_\alpha^+(t) + j_\alpha^-(t)$, $j_\alpha^{\pm}(t) = j_{0\alpha}^{\pm}(t) + j_{1\alpha}^{\pm}(t)$, where

$$j_\alpha^{\pm}(t) = \frac{-2e}{(2\pi)^2} \int_{BZ} v_\alpha^{\pm}(t,\mathbf{q}) f^{\pm}(t,\mathbf{q}) d^2 q + \frac{-2e^2 \eta_0}{(2\pi)^2 c} \int_{BZ} [m_{\alpha\beta}^{\pm}(t,\mathbf{q})]^{-1} A_\beta(t) f^{\pm}(t,\mathbf{q}) d^2 q. \qquad (5)$$

Here, the index $\nu=0$ corresponds to the zero degree of the vector potential, and $\nu=1$ corresponds to the first degree, i.e., respectively, to the first and second terms in (5). Also here $v_\alpha^{\pm}(\mathbf{q}) = \pm(\gamma_0/\hbar)\partial_{q_\alpha} w(\mathbf{q})$ is the velocity components, $[m_{\alpha\beta}^{\pm}(\mathbf{q})]^{-1} = \pm(\gamma_0/\hbar^2)\partial_{q_\alpha}\partial_{q_\beta} w(\mathbf{q})$, and $m_{\alpha\beta}$ is the component of the mass tensor. For electrons DF is $f_e(t,\mathbf{q}) = f^+(t,\mathbf{q})$, for holes $f_h(t,\mathbf{q}) = f^-(t,\mathbf{q}) = 1 - f^+(t,\mathbf{q})$. In the zero approximation $f^{\pm}(t,\mathbf{q}) = [1 + \exp(\pm(\gamma_0 w(t,\mathbf{q}) - \mu_c))]^{-1}$, i.e. it is a Fermi-Dirac function (FDF) of the perturbed energy $\gamma_0 w(t,\mathbf{q})$ in the zones. For physical reasons, the tensor $m_{\alpha\beta}$ is diagonal. In (3), we took into account the spin degeneracy $g_s = 2$, $(t,\mathbf{q})$ means the time dependence of $\mathbf{q}$. In our case, the equation (5) includes the effective mass component

$$m_{xx}^{\pm}(t,\mathbf{q}) = \frac{\mp \hbar^2 w^3(t,\mathbf{q})/(3\gamma_0 a^2)}{\cos\left(\frac{q_y a}{2}\right)\left[\cos\left(\frac{q_y a}{2}\right)\sin^2\left(\sqrt{3}a\frac{q_x - q_0(t)}{2}\right) + \frac{w^2}{2}(\mathbf{q})\cos\left(\sqrt{3}a\frac{q_x - q_0(t)}{2}\right)\right]}, \quad (6)$$

as well as the speed component

$$v_x^{\pm}(t,\mathbf{q}) = \pm \frac{\gamma_0}{\hbar} w'_{q_x}(t,\mathbf{q}) = \mp \frac{2v_F}{w(t,\mathbf{q})}\sin\left(\sqrt{3}a\frac{q_x - q_0(t)}{2}\right)\cos\left(a\frac{q_y}{2}\right). \quad (7)$$

Here we have taken into account the perturbation of these quantities under the action of a vector potential, which changes the quasi-pulse $p_x$ to $p_x - e\eta_0 A_x/c$, or $q_x \to q_x - q_0(t)$, $q_0(t) = e\eta_0 A_x(t)/(\hbar c)$. Without taking this perturbation into account, the linear contribution is given only by the second term in (5), and the first one vanishes. As a result of the perturbation, functions (6) and (7) become functions of time. The zone structure (1) becomes the same function of time, in which $w(t,\mathbf{q}) = w(q_x - q_0(t), q_y)$. The dependence of the normalized velocity component $v_x^-(t,\mathbf{q})/v_F$ of electron in $\pi^-$ zone is shown in Fig. 4.

## 3. Nonstationary equations

Let the wave front approach graphene at the moment $t=0$ and the excitation lasts for a very long time. The incident field (3) exists on graphene only at $t \geq 0$. In the full field, the graphene effect itself should be taken into account, i.e., the generation of waves by its current density. At time $t$, the vector potential of the diffraction field is

$$A_x^d(z,t) = \frac{c}{2}\int_0^{t-|z|/c} j_x(t')dt', \quad (8)$$

in this case, the field on graphene is $E_x^d(0,t) = -\eta_0 \partial_t A_x^d(0,t)/c = -\eta_0 j_x(t)/2$, $H_y^d(\pm 0,t) = \partial_z A_x^d(0,t) = -\mathrm{sgn}(z)j_x(0)/2$, where $j_x(t) = H_y^d(-0,t) - H_y^d(+0,t)$. If the current at large $t$ contains the harmonics, then the diffraction field also contains the same harmonics (or the same spectrum). The full field satisfies the equation

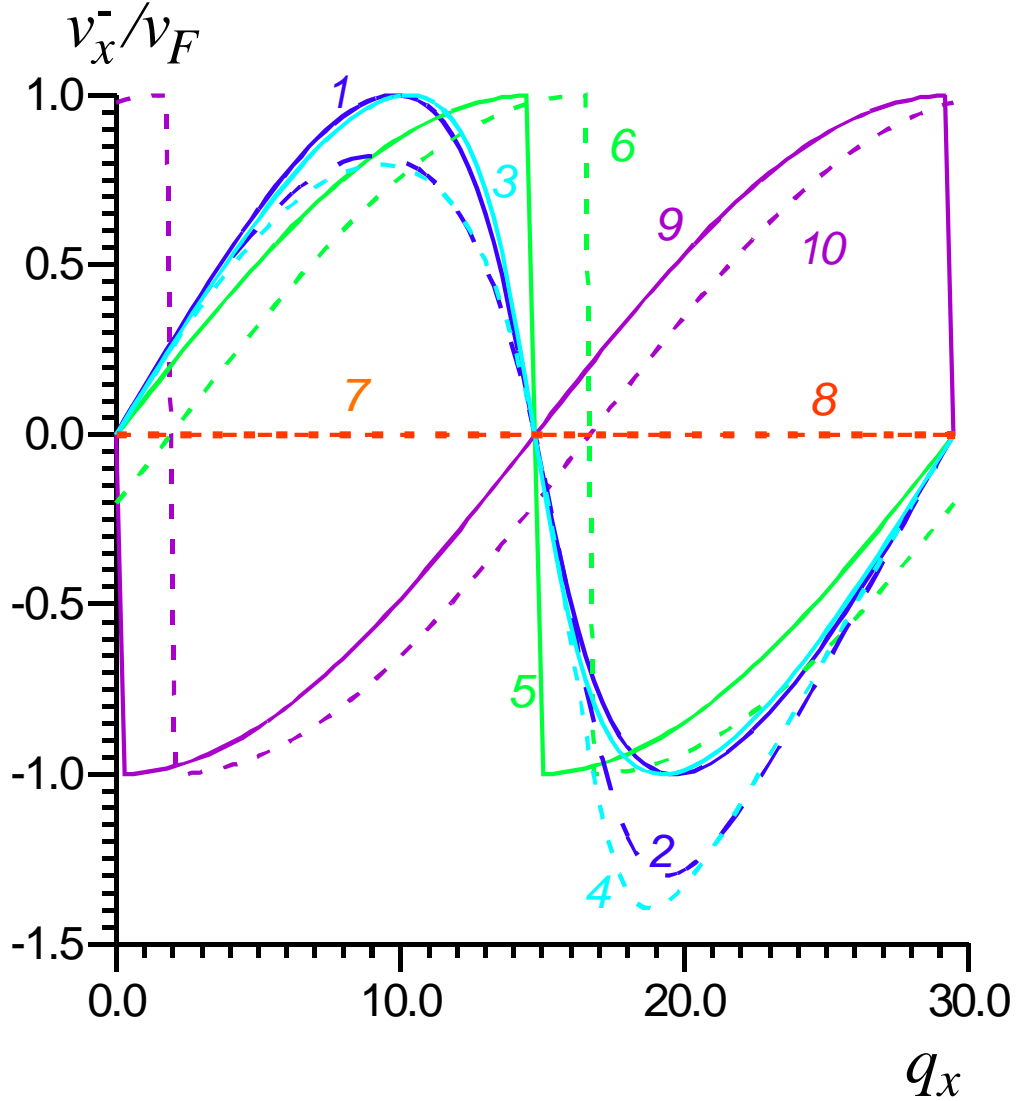

Fig. 4. Dependence of the normalized velocity component $v_x^-(\mathbf{q})/v_F$ (solid lines) and $v_x^-(t,\mathbf{q})/v_F$ (dashed lines) from $q_x$ (1/nm) at $q_0=1.91$ nm$^{-1}$ ($A_x$=1A) and different $q_y$: $q_y = 0$ (curves 1), $q_y = \pi/(3a)$ (2), $q_y = 2\pi/(3a)$ (3), $q_y = \pi/a$ (4), $q_y = 4\pi/(3a)$ (5)

$$E_x(0,t,k_x) = E_0 \tilde{k} \cos(\omega_0 t) - \frac{\eta_0}{2}\int_0^t \Sigma_{xx}(t-t',k_x)E_x(0,t',k_x)dt'. \qquad (9)$$

Here $\Sigma_{xx}(t,k_x)$ is a certain operator, generally speaking, integral and nonlinear in $E_0$. The integral in (9) describes the total current $j_x(t,k_x)$. Since all quantities are determined at a large point in time, their spectra are instantaneous, i.e. they depend on frequency and time, but quasi-monochromatic. The current determined by the conductivity has the form

$$j_{1x}(t,k_x) = \frac{1}{2\pi} \int_{-\infty}^{\infty} \exp(-i\omega t) \sigma_{xx}(t,\omega,k_x) E_x(t,\omega,k_x) d\omega, \tag{10}$$

$\sigma_{xx}(t,\omega,k_x)$ is the instantaneous spectrum of total conductivity which generally defines an integral operator of the form included in (9)

$$j_{1x}(t,k_x) = \int_0^t \Sigma_{xx}(t-t',k_x) E_x(0,t',k_x) dt'.$$

This leads to the need to solve nonlinear equations. In the approximation of an incident field at high times $E_x(\omega) = -i\omega\eta_0 A_x(\omega)/c$. Thus, (5) gives a nonlinear response. The absence of a field leads to the fact that the first term in (5), when integrated over the BZ, vanishes, as does the second. This is due to the fact that $f^{\pm}(0,\mathbf{q})$ are even functions of the variable $q_x$, and $v_x^{\pm}(0,\mathbf{q})$ are odd functions. Integrating the integral in parts in the second term in (5), we find

$$\frac{2e^2\eta_0}{(2\pi)^2 ck_B T} \int_{BZ} A_x(t) v_x^{\pm 2}(t,\mathbf{q}) f^{\pm}(t,\mathbf{q})[1 - f^{\pm}(t,\mathbf{q})] d^2q. \tag{11}$$

When transferring the operator $\partial_{q_x}$ from $v_x^{\pm}$ to $f^{\pm}(t,\mathbf{q})$ under the substitutions $q_x = \pm 2\pi/(\sqrt{3}a)$ in the integration over the rectangular part of the BZ (Fig. 1) we have zero due to periodicity. What remains is the integral corresponding to the triangular part

$$2\int_{2\pi/(3a)}^{4\pi/(3a)} \left[ v_x^{\pm}(t,q_x,q_y) f^{\pm}(t,q_x,q_y) - v_x^{\pm}(t,-q_x,q_y) f^{\pm}(t,-q_x,q_y) \right] dq_y.$$

We have $v_x^{\pm}(t,\pm q_x,q_y) = 0$ at $q_y = \pi/a$ (7). By making the substitution $q_y' = q_y - \pi/a$, we get the integral

$$2\int_{-\pi/(3a)}^{\pi/(3a)} v_x^{\pm}(t,q_x,q_y') \left[ f^{\pm}(t,q_x,q_y') + f^{\pm}(t,-q_x,q_y') \right] dq_y'.$$

It has an odd function $v_x^{\pm}(t,q_x,q_y')$ of $q_y'$, and $q_x$ is the function of $q_y'$. At $t=0$, the function in parentheses is even, so the integral is zero. When using decompositions by $q_0$, it is also easy to prove this statement. Thus, the second part of formula (5) is given by the integral (11). The complex reaction to prolonged exposure to the field (3) has the form of a current

$$\dot{j}_{1x}(t) = \dot{\sigma}_{xx1}(t)E_0\tilde{k}\exp(-i\omega_0 t) + \dot{\sigma}_{xx3}(t)(E_0\tilde{k})^3\exp(-3i\omega_0 t) + ... \tag{12}$$

Complex amplitudes are used here, i.e. the real current should be taken as the real part: $j_{1x}(t) = \text{Re}(\dot{j}_{1x}(t))$.. The field spectrum on graphene at positive frequencies has the form

$$E_x(t,\omega,k_x) \approx E_0\tilde{k}\delta(\omega-\omega_0) + E_0\tilde{k}\kappa_{13}\delta(\omega-3\omega_0) + ... + \delta E_x(t,\omega,k_x), \tag{13}$$

and at negative frequencies, it is complex conjugate to (13). The component $\delta E_x(t,\omega,k_x)$ is lower the longer the exposure, i.e., the greater the time $t$. Here is the coefficient of transformation of the first harmonic into the third harmonic in the field. The spectral response has the form $j_{1x}(t,\omega) = \sigma_{xx}(t,\omega,k_x)E_x(t,\omega,k_x)$. Substituting (13) into (12), we obtain

$$j_{1x}(t,k_x) = \frac{1}{\pi}\text{Re}\int_0^\infty \exp(-i\omega t)\sigma_{xx}(t,\omega,k_x)E_x(t,\omega,k_x)d\omega = $$
$$2\text{Re}\left[\exp(-i\omega_0 t)\sigma_{xx}(t,\omega_0,k_x)E_0\tilde{k} + \exp(-3i\omega_0 t)\sigma_{xx}(t,3\omega_0,k_x)E_0\tilde{k}\kappa_{13}(3\omega_0)\right] \tag{14}$$

Because $H_y^d(\pm 0,t) = \pm E_x^d(0,t)/\eta_0$ we have $j(t) = 2E_x^d(0,t)/\eta_0$, which corresponds to the previously obtained condition. Since $\sigma_{xx}$ depends on the vector potential and, consequently, on the field amplitude $E_0$, (14) can be written as

$$j_{1x}(t,k_x) = \sigma_{1c}(t,\tilde{k})\cos(\omega_0 t)E_0 + \sigma_{1s}(t,\tilde{k})\sin(\omega_0 t)E_0 + $$
$$+ \sigma_{3c}(t,\tilde{k})\cos(3\omega_0 t)E_0^3 + \sigma_{3s}(t,\tilde{k})\sin(3\omega_0 t)E_0^3 + ..., \tag{15}$$

and also in the form

$$j_{1x}(t,k_x) = [\sigma'_{xx1}(t)\cos(\omega_0 t) + \sigma''_{xx1}(t)\sin(\omega_0 t)]E_0\tilde{k} + $$
$$+ [\sigma'_{xx3}(t)\cos(3\omega_0 t) + \sigma''_{xx3}(t)\sin(3\omega_0 t)](E_0\tilde{k})^3 + ...$$

Accordingly, we have

$$\sigma_{1c}(t,\tilde{k}) = \sigma'_{xx}(t,\omega_0,k_x)\tilde{k},$$
$$\sigma_{1s}(t,\tilde{k}) = -\sigma''_{xx}(t,\omega_0,k_x)\tilde{k}, \tag{16}$$
$$\sigma_{3c}(t,\tilde{k}) = [\sigma'_{xx}(t,3\omega_0,k_x)k'_{13}(3\omega_0) - \sigma''_{xx}(t,3\omega_0,k_x)k''_{13}(3\omega_0)]\tilde{k}/E_0^2,$$
$$\sigma_{3c}(t,\tilde{k}) = [\sigma'_{xx}(t,3\omega_0,k_x)k''_{13}(3\omega_0) + \sigma''_{xx}(t,3\omega_0,k_x)k'_{13}(3\omega_0)]\tilde{k}/E_0^2.$$

Also $\sigma'_{xx1}(t) = \sigma_{1c}(t,\tilde{k})/\tilde{k}$, $\sigma''_{xx1}(t) = \sigma_{1s}(t,\tilde{k})/\tilde{k}$, $\sigma'_{xx3}(t) = \sigma_{3c}(t,\tilde{k})/\tilde{k}^3$, $\sigma''_{xx3}(t) = \sigma_{3s}(t,\tilde{k})/\tilde{k}^3$. Here, a stroke means real parts, and a double stroke means imaginary parts. To determine the corresponding (4) operator $\Sigma_{xx}(t) = \Sigma_{xx}^{intra}(t) + \Sigma_{xx}^{inter}(t)$, (4) should be reversed in Fourier, while taking the complex frequency $\omega + i\omega_c$ and considering the poles $\omega_{pn}(t)$ and $\pm\Omega(t)$ in the lower half-plane of the complex frequency plane. Integration by the method of deductions leads to attenuating time dependencies $\exp(-\omega_c t)$. We denote the energies $\varepsilon_n(t, \mathbf{p}\pm) = (-1)^n \gamma_0 w(q_x - q_0(t) \pm k_x/2)$ by the function $w(q_x, q_y)$ (1). Then the integrals take place for the intraband conductivity

$$\frac{1}{2\pi} \int_{-\infty}^{\infty} \frac{\exp(-i\omega t)}{\omega - \omega_{pn}} d\omega = -i\exp(-i\omega_{pn} t),$$

in which $n = 1,2$ and the frequencies $\omega_{pn} = (-1)^n \tilde{\omega}_p(t, \mathbf{q}) + i\omega_c$, $\tilde{\omega}_p(t, \mathbf{q}) = \gamma_0 [w(q_x - q_0(t) + k_x/2, q_y) - w(q_x - q_0(t) - k_x/2, q_y)]/\hbar$ are indicated. In the result we have

$$\Sigma_{xx}^{\text{intra}}(t) = \frac{-2e^2}{\pi^2 \hbar} \sum_{n=1,2} \int_{BZ} \frac{d^2q \, v_x^2(t,\mathbf{q})\{f_0(\varepsilon_n(t,\mathbf{p}^-)) - f_0(\varepsilon_n(t,\mathbf{p}^+))\}}{\tilde{\omega}_p(t,\mathbf{q})} \exp(-\omega_c t)\cos(\tilde{\omega}_p(t,\mathbf{q})t). \quad (17)$$

For interband conductivity, we introduce the frequency square $\Omega^2(t,\mathbf{q}) = [w(q_x - q_0(t) + k_x/2, q_y) + w(q_x - q_0(t) - k_x/2, q_y)]^2 / \hbar^2$. This frequency depends on the time. A dimensionless integral arises:

$$I(t,\mathbf{q}) = \frac{i}{2\pi} \int_{-\infty}^{\infty} \frac{\omega \exp(-i\omega t) d\omega}{(\omega + i\omega_c)^2 - \Omega^2} =$$

$$= \frac{i}{4\pi\Omega} \int_{-\infty}^{\infty} \left[\frac{\omega}{\omega + i\omega_c - \Omega} - \frac{\omega}{\omega + i\omega_c + \Omega}\right] \exp(-i\omega t) d\omega \quad (18)$$

We have added a small imaginary part to the frequency in the denominator (18) $\omega \to \omega + i\omega_c$, otherwise the integral of the form (18) diverges. This can be explained by the fact that function (4) must have the property $\sigma_{\alpha\beta}(-\omega,\mathbf{k}) = \sigma_{\alpha\beta}^*(\omega,\mathbf{k})$ for a real time response. The integral (18) under the specified substitution has this property. Integrating it by the method of deductions, we obtain

$$I(t,\mathbf{q}) = \frac{\Omega(t,\mathbf{q})\cos(\Omega(t,\mathbf{q})t) - \omega_c \sin(\Omega(t,\mathbf{q})t)}{\Omega(t,\mathbf{q})} \exp(-\omega_c t). \quad (19)$$

We have the result

$$\Sigma_{xx}^{\text{inter}}(t) = \frac{e^2}{\pi^2 \hbar} \int_{BZ} \frac{I(t,\mathbf{q})}{\Omega(t,\mathbf{q})} \{f_0(\varepsilon_1(t,\mathbf{p}^-)) - f_0(\varepsilon_2(t,\mathbf{p}^+))\} v_x^2(t,\mathbf{q}) d^2q. \quad (20)$$

It should be noted that relaxation times (and collision frequencies $\omega_c$) may depend on energies or impulses, and generally speaking, they are different for interband transitions than for intraband ones. (17) and (20) include the velocity (7). Figure 4 shows the results of calculating the velocity $v_x^-(t,\mathbf{q})$ of an electron in the VZ. Since the conversion coefficient to the third harmonic is small, and in the first approximation, with a normal wave incidence, we have an oscillating value $q_0 = eE_0 \sin(\omega_0 t)/(\hbar\omega_0)$, then taking its maximum $eE_0/(\hbar\omega_0)$ and assuming it to be equal to the average value of $q_x$ in the order of magnitude $\pi/a$, we obtain a critical electric field at which the distortion of the BZ by the wave can be considered strong. We have $E_c = \hbar\omega_0\pi/(ea)$. For the $\omega_0 = 10^{13}$ Hz frequency, we get $E_c = 8.4 \cdot 10^7$ V/m. For fields an order of magnitude smaller, the distortion of the BZ can already be considered small. Note that the energy density in the field with $E_0 = 10^6$ V/m is 4.4 J/m$^3$, and for the critical field it is almost $10^4$ times higher. Now it is possible to solve the nonlinear equation (9). The simplest way to get a solution is to use the second approximation, substituting the first one $E_0 \tilde{k} \cos(\omega_0 t)$ into the integral. The result of the integration should be reversed by Fourier and the harmonic contribution calculated. It is also possible to obtain the third and subsequent approximations, but this requires calculating multiple integrals. A simpler and more accurate method is to integrate (9) numerically using small sampling areas $\Delta t$ of the domain. We denote the field $E_{x0} = E_0 \tilde{k}$ at the moment $t = 0$. The field at the moment is calculated as

$$E_{x1} = E_0 \tilde{k} \cos(\omega_0 \Delta t) - \frac{\eta_0}{2} E_{x1} \int_0^{\Delta t} \Sigma_{xx}(\Delta t, k_x) dt' = E_0 \tilde{k}\left(\cos(\omega_0 \Delta t) - \frac{\eta_0}{2}\Sigma_{xx}(\Delta t, k_x)\Delta t\right). \qquad (21)$$

The subsequent value $E_{x2}$ is calculated using $E_{x1}$:

$$E_{x2} = E_0 \tilde{k} \cos(2\omega_0 \Delta t) - \frac{\eta_0}{2} E_{x1} \int_{\Delta t}^{2\Delta t} \Sigma_{xx}(\Delta t, k_x) dt' =$$
$$= E_0 \tilde{k} \cos(2\omega_0 \Delta t) - \frac{\eta_0}{2} E_{x1} \Sigma_{xx}(\Delta t, k_x)\Delta t.$$

More precise quadrature formulas for calculating integrals can be also used. Denoting $e_x(t) = E_x(0, t, k_x)$ and differentiating (9), we obtain the integro-differential equation

$$e'_x(t) = -\omega_0 E_0 \tilde{k} \sin(\omega_0 t) - \frac{\eta_0}{2} \int_0^t \Sigma'_{xx}(t - t', k_x) e_x(t') dt' - \frac{\eta_0}{2} \Sigma_{xx}(0, k_x) e_x(t). \qquad (22)$$

By solving the Cauchy problem (22), it is also possible to obtain a periodogram of the electric field. In this equation, the functions $\Sigma'_{xx}(t - t', k_x)$ and $\Sigma_{xx}(0, k_x)$ depend on the field $e_x(t)$, which is not explicitly stated.

## 4. The perturbation method

We will use parameter $q_0(t)$ expansions up to the third order for (1)
$w(t,\mathbf{q}) = w(\mathbf{q}) + w_1(\mathbf{q})q_0(t) + w_2(\mathbf{q})q_0^2(t) + ...$, velocity $v_x^\pm(t,\mathbf{q}) = \pm v_F\left(v_0(\mathbf{q}) + v_1(\mathbf{q})q_0(t) + v_2(\mathbf{q})q_0^2(t) + ...\right)$, and DF $f^\pm(t,\mathbf{q}) = f^\pm(\mathbf{q}) + f_1^\pm(\mathbf{q})q_0(t) + f_2^\pm(\mathbf{q})q_0^2(t) + ...$. All coefficients are given in Appendix (A1)–(A3). Consider the normal wave incidence when $k_x = 0$, and there is no spatial dispersion. In this case $\tilde{\omega}_p(t) \to 0$, and $\Omega(t,\mathbf{q}) = 2\gamma_0 w(t,\mathbf{q})/\hbar$. We have decomposition $w(q_x - q_0, q_y) \approx w(\mathbf{q}) + w_1(\mathbf{q})q_0 + w_2(\mathbf{q})q_0^2$. The function $w_1$ is odd in $q_x$, and the function $w_2$ is even one. We 'll find $w(q_x - q_0(t) \pm k_x/2, q_y) \approx w(t,\mathbf{q}) \pm w_1(t,\mathbf{q})k_x/2$ for a small $k_x$. For Fermi-Dirac DF

$$f_0(\varepsilon_n(t,\mathbf{p}\pm)) = f_0(\varepsilon_n(t,\mathbf{p})) - (-1)^n f_0^2(\varepsilon_n(t,\mathbf{p}))\frac{\gamma_0 \, w_1(t,\mathbf{q})}{k_B T}\left(\pm \frac{k_x}{2}\right) + ... \quad (23)$$

Now $f_0(\varepsilon_n(t,\mathbf{p}^-)) - f_0(\varepsilon_n(t,\mathbf{p}^+)) = (-1)^n f_0^2(\varepsilon_n(t,\mathbf{p}))\frac{\gamma_0 w_1(t,\mathbf{q})}{k_B T} k_x$, and we have

$$\Sigma_{xx}^{\text{intra}}(t) = \frac{-4e^2}{\pi^2 k_B T}\exp(-\omega_c t)\int_{BZ} v_x^2(t,\mathbf{q})d^2q. \quad (24)$$

For interband conductivity

$$\Sigma_{xx}^{\text{inter}}(t) = \frac{e^2}{2\pi^2 \gamma_0}\int_{BZ}\frac{I(t,\mathbf{q})}{w(t,\mathbf{q})}\{f_0(-\gamma_0 w(t,\mathbf{q})) - f_0(\gamma_0 w(t,\mathbf{q}))\}v_x^2(t,\mathbf{q})d^2q. \quad (25)$$

All functions depend only on $w(t,\mathbf{q})$. Decomposing the latter into a series of coefficients (A1) and counting $q_0(t)$ as a small value, we obtain

$$\Sigma_{xx}^{\text{intra}}(t) = \Sigma_{xx0}^{\text{intra}}(t) + \Sigma_{xx2}^{\text{intra}}(t)q_0^2(t) + ..., \quad (26)$$

$$\Sigma_{xx}^{\text{inter}}(t) = \Sigma_{xx0}^{\text{inter}}(t) + \Sigma_{xx2}^{\text{inter}}(t)q_0^2(t) + .... \quad (27)$$

(26) and (27) include the values

$$\Sigma_{xx0}^{\text{intra}}(t) = \frac{-4e^2}{\pi^2 k_B T}\exp(-\omega_c t)\int_{BZ} v_x^2(\mathbf{q})d^2q,$$

$$\Sigma_{xx0}^{\text{inter}}(t) = \frac{e^2}{2\pi^2 \gamma_0}\int_{BZ}\frac{I(\mathbf{q})}{w(\mathbf{q})}\{f_0(-\gamma_0 w(\mathbf{q})) - f_0(\gamma_0 w(\mathbf{q}))\}v_x^2(\mathbf{q})d^2q,$$

determining the linear response. In them the function $w(\mathbf{q})$ is determined by equation (1), $v_x^2(\mathbf{q})$ is obtained by squaring (21) at $q_0 = 0$, the integral $I(\mathbf{q})$ is given by the formula (19) at $q_0 = 0$.

Decompositions (26) and (27) include linear terms (and generally odd terms). For them, coefficients are odd functions of the variable $q_x$, and for them the integral over BZ is zero. Therefore, odd harmonics occur during integration. To obtain the transformation coefficients $\Sigma_{xx2}^{intra}(t)$, it is sufficient to obtain the decomposition of the velocity (7) in $q_0$. It has the form (A2). The functions $v_0$ and $v_2$ are odd in $q_x$, and the functions $v_1$ and $v_3$ are even. All functions are even in $q_y$. When squaring the velocity, we omit the odd terms and get

$$\Sigma_{xx2}^{intra}(t) = \frac{-4e^2 v_F^2}{\pi^2 k_B T} \exp(-\omega_c t) \int_{BZ} \left(v_1^2(\mathbf{q}) + 2v_0(\mathbf{q})v_2(\mathbf{q})\right) d^2q. \quad (28)$$

In the approximation of an incident wave $E_0 \cos(\omega_0 t)$, it is easy to obtain a third harmonic. As for the interband conductivity, the decompositions $I(t,\mathbf{q}) = I(\mathbf{q}) + I_1(\mathbf{q})q_0 + I_2(\mathbf{q})q_0^2 + ...$ and $f_0(\pm\gamma_0 w(t,\mathbf{q})) = f_0(\pm\gamma_0 w(\mathbf{q})) + f_1^\pm(\mathbf{q})q_0 + f_2^\pm(\mathbf{q})q_0^2 +$ should be used for it. As a result, we get

$$\Sigma_{xx2}^{inter}(t) = \frac{e^2 v_F^2}{2\pi^2 \gamma_0} q_0^2(t) \int_{BZ} F(\mathbf{q}) d^2q, \quad (29)$$

where we have a rather cumbersome expression

$$F(\mathbf{q}) = \left(f_{e2}^-(\mathbf{q}) - f_{e2}^+(\mathbf{q})\right)\frac{I(\mathbf{q})}{w(\mathbf{q})} v_0^2(\mathbf{q}) + 2v_0(\mathbf{q})v_1(\mathbf{q})\left(f_0^-(\mathbf{q}) - f_0^+(\mathbf{q})\right)\left[\frac{I_1(\mathbf{q})q_0}{w(\mathbf{q})} - \frac{I(\mathbf{q})w_1(\mathbf{q})}{w^2(\mathbf{q})}\right] +$$
$$+ \left(f_1^-(\mathbf{q}) - f_1^+(\mathbf{q})\right)\left(\frac{I_1(\mathbf{q})}{w(\mathbf{q})}v_0^2(\mathbf{q}) - \frac{I(\mathbf{q})w_1(\mathbf{q})}{w^2(\mathbf{q})} + 2v_0(\mathbf{q})v_1(\mathbf{q})\frac{I(\mathbf{q})}{w(\mathbf{q})}\right) +$$
$$+ \left(f_0^-(\mathbf{q}) - f_0^+(\mathbf{q})\right)\left(v_0^2(\mathbf{q})\left[I(\mathbf{q})\frac{w_1^2(\mathbf{q}) - w(\mathbf{q})w_2(\mathbf{q})}{w^3(\mathbf{q})} - \frac{I_1(\mathbf{q})w_1(\mathbf{q})}{w^2(\mathbf{q})} + \frac{I_2(\mathbf{q})}{w(\mathbf{q})}\right] + \frac{I(\mathbf{q})}{w(\mathbf{q})}v_1^2(\mathbf{q})\right)$$

The functions $f_n^\pm(\mathbf{q})$ are defined by formulas (A3). You can also get a fifth-order contribution, but the calculations become very cumbersome. However, we did not take into account the contribution from the first member in (5). Its contribution to the first and third harmonics has the form

$$j_{0x}^\pm(t) \mp \frac{2ev_F q_0(t)}{(2\pi)^2} \tilde{F}_1(\mu_c, T) \mp \frac{2ev_F q_0^3(t)}{(2\pi)^2} \tilde{F}_3(\mu_c, T), \quad (30)$$

Where $\tilde{F}_{1,3}(\mu_c, T) = \int_{BZ} F_{1,3}(\mathbf{q}) d^2q$ are the integrals over BZ, and integral functions are defined as

$$F_1(\mathbf{q}) = v_0(\mathbf{q})f_1^\pm(\mathbf{q}) + v_1(\mathbf{q})f_0^\pm(\mathbf{q}), \qquad F_3(\mathbf{q}) = \left(v_0(\mathbf{q})f_3^\pm(\mathbf{q}) + v_1(\mathbf{q})f_2^\pm(\mathbf{q}) + v_2(\mathbf{q})f_1^\pm(\mathbf{q}) + v_3(\mathbf{q})f_0^\pm(\mathbf{q})\right).$$

Also, instead of using Fourier inversion (4), we can determine the contribution from the second term to (5). These results give time processes that should be reversed in Fourier order to obtain harmonics. With prolonged exposure, it is possible to switch to spectral amplitudes, considering the processes as monochromatic and making a substitution $q_0(t) = \tilde{q}_0 \exp(-i\omega_0 t)$. This is possible at small field amplitudes, when there is no heating and the chemical potential is unchanged. In our case for spectral amplitudes $\eta_0 A_x(\omega)/c = i\omega E_x(\omega)$.

Substitution of a vector potential in (1) is not a completely justified procedure in a strong field. In reality, the value $-e\eta_0 A_x(t)$ should be substituted into SE, which becomes non-stationary with a periodic graphene potential [14]. The energy is not defined in such case. However, the wave function is not defined for it either, since the system is not closed. Graphene is in thermodynamic equilibrium with the Planck thermal field and is a subsystem interacting with a large reservoir (thermostat). Therefore, the subsystem does not have a definite wave function, and its mixed states are described using a density matrix that replaces the wave function [53]. In our case of the periodic effects the DF $f^\pm$ can be considered as diagonal elements of the density matrix satisfying the kinetic Boltzmann equation (KBE) in the form of BGK. It is convenient to consider it in a complex form. Then we have the decomposition $f^\pm(t,\mathbf{q}) = f_1^\pm(\mathbf{q})\exp(-i\omega_0 t) + f_3^\pm(\mathbf{q})\exp(-3i\omega_0 t)$ for DF, and we represent the field in terms of odd harmonics $E_x = E_0 E_1 \exp(-i\omega_0 t) + E_0 E_3 \exp(-3i\omega_0 t)$. The complex KBE has the form

$$\left(i\omega_0 f_1^\pm + \frac{eE_0 E_1}{\hbar}\nabla_{q_x} f_1^\pm(\mathbf{q})\right)\exp(-i\omega_0 t) + \left(3i\omega_0 f_3^\pm + \frac{eE_0 E_3}{\hbar}\nabla_{q_x} f_3^\pm(\mathbf{q})\right)\exp(-3i\omega_0 t)$$
$$= \frac{f_1^\pm(\mathbf{q})\exp(-i\omega_0 t) + f_3^\pm(\mathbf{q})\exp(-3i\omega_0 t) - f_0^\pm(\mathbf{q})\exp(-i\omega_0 t)}{\tau_r}. \quad (31)$$

Equation (31) splits into two:

$$(1 - i\omega_0 \tau_r) f_1^\pm(\mathbf{q}) = f_0^\pm(\mathbf{q}) + \frac{eE_0 E_1 \tau_r}{\hbar}\nabla_{q_x} f_1^\pm(\mathbf{q}), \quad (32)$$

$$(1 - 3i\omega_0 \tau_r) f_3^\pm(\mathbf{q}) = \frac{eE_0 E_3 \tau_r}{\hbar}\nabla_{q_x} f_3^\pm(\mathbf{q}). \quad (33)$$

We use the approximation $\nabla_{q_x} f^\pm(\mathbf{q}) \approx \nabla_{q_x} f_0^\pm(\mathbf{q}) = \frac{\hbar v_x^\pm}{k_B T}\exp(\pm u(\mathbf{q})) f_0^\pm(\mathbf{q})[1 - f_0^\pm(\mathbf{q})]$. Here $u(\mathbf{q}) = \gamma_0 w(\mathbf{q}) - \mu_c$. Then we get the solutions

$$f_1^\pm(\mathbf{q}) = \frac{f_0^\pm(\mathbf{q}) + \frac{eE_0 E_1 \tau_r v_x^\pm(\mathbf{q})}{k_B T}\exp(\pm u(\mathbf{q})) f_0^\pm(\mathbf{q})[1 - f_0^\pm(\mathbf{q})]}{1 - i\omega_0 \tau_r}, \quad (34)$$

$$f_3^\pm(\mathbf{q}) = \frac{\dfrac{eE_0 E_3 \tau_r v_x^\pm(\mathbf{q})}{k_B T} \exp(\pm u(\mathbf{q})) f_0^\pm(\mathbf{q})[1 - f_0^\pm(\mathbf{q})]}{1 - 3i\omega_0 \tau_r}. \tag{35}$$

In them, the magnitudes $E_1$ and $E_3$ are complex and $E_3$ determine the generation of the third harmonic, but it depends on the DF. The deviation in (35) from FDF will be and small at $eE_0 \tau_r v_F \ll k_B T$. Assuming the relaxation time $\tau_r = 10^{-13}$ s, at room temperature we get $E_0 \ll 10^{10}$ V/m. Therefore, with not too strong fields and not too low temperatures, it is quite possible to take $f^\pm(\mathbf{q}) = f_0^\pm(\mathbf{q})$. Using (34) and (35) greatly complicate the result. Such a DF should be considered as $f^\pm(t, \mathbf{q}) = \text{Re}(f_1^\pm(\mathbf{q})\exp(-i\omega_0 t) + f_3^\pm(\mathbf{q})\exp(-3i\omega_0 t))$. Considering in (32) and (33) the DF $f_0^\pm(t, \mathbf{q})$ and velocities as perturbed quantities decomposed by $q_0(t) = \tilde{q}_0 \exp(-i\omega_0 t)$, more complex relations can be obtained. There are unknowns $E_1$, $E_3$ in the considered approach, which should be determined iteratively, since the problem is nonlinear. You should take $\tilde{q}_0 = -eE_0 i/(\omega_0 \hbar)$. Then $\text{Re}(q_0(t)) = -eE_0 \sin(\omega_0 t)/(\omega_0 \hbar)$, and the real electric field of the incident wave is $E_x = E_0 \cos(\omega_0 t)$. Note that if we set the DF in the center of the BZ (for example, in the form of FDF), then equations (32) and (33) can be considered as Cauchy problems.

Thus, a rigorous approach requires calculating for each moment $t$ the response (5) while simultaneously solving the KBE in the time domain and the problem of excitation of the scattered field by current. The corresponding solutions should be used in (5). To obtain the harmonic composition, the instantaneous spectra should be calculated for a given time. This approach is very costly: integration over the BZ should be carried out for each moment in time, and there are a lot of such moments for the quasi-monochromatic process. The integrals over BZ are not analytically calculated. Therefore, to get a response, it is easier to use decompositions to integrate once. It is possible to integrate according to BZ in Fig. 1 by performing double integrations over triangles with a variable limit. If $I_1(A(t))$ is the integral of triangle 1 in Fig. 1, then the integral $I_1(-A(t))$ of the triangle mirrored from the y axis should be added to it, as well as the integrals $I_2(A(t))$, $I_2(-A(t))$, $I_3(A(t))$, $I_3(-A(t))$, corresponding to triangles 2 and 3. Due to the parity in $q_y$, the result should be doubled, which corresponds to the triangular areas in the lower half of the BZ. It is also possible to integrate over a rectangular area and over an upper triangle, similar to obtaining the ratio (11). The easiest way to integrate is by shooting in a double loop over the areas $-2\pi/(\sqrt{3}a) < q_x < 2\pi/(\sqrt{3}a)$ and $0 < q_y < 4\pi/(3a)$, taking into account only those points that hit the BZ, i.e. for which $q_y < 4\pi/3a - |q_x|/\sqrt{3}$. The result should be doubled. The use of decompositions makes it easier to obtain a response in the form of a third harmonic. Let the field not be too strong, and the change of DF into a KBE can be ignored, i.e. we take into account only the expansion (A3). We also do not take into account the change $A_x$ due to the action of the current. It is also possible to take into account the decomposition of the mass in (5). It is more convenient to take (5) taking into account (11) in the form

$$j_x^{\pm}(t) = \frac{2e}{(2\pi)^2} \int_{BZ} v_x^{\pm}(t,\mathbf{q}) f^{\pm}(t,\mathbf{q}) \left[ \frac{q_0(t) v_x^{\pm} \hbar}{k_B T} (1 - f^{\pm}(t,\mathbf{q})) - 1 \right] d^2q.$$

We will rewrite this expression in the form $j_x^{\pm}(t) = j_{x1}^{\pm} q_0(t) + j_{x3}^{\pm} q_0^3(t)$, where

$$j_{x1,3}^{\pm}(t) = \frac{1}{(2\pi)^2} \int_{BZ} F_{1,3}^{\pm}(\mathbf{q}) d^2q,$$

$$F_1^{\pm}(\mathbf{q}) = \mp 2ev_F \left( v_0(\mathbf{q}) f_1^{\pm}(\mathbf{q}) + v_1(\mathbf{q}) f_0^{\pm}(\mathbf{q}) - v_0(\mathbf{q}) \frac{v_x^{\pm} \hbar}{k_B T} f_0^{\pm}(\mathbf{q})(1 - f_0^{\pm}(\mathbf{q})) \right),$$

$$F_3^{\pm}(\mathbf{q}) = \mp 2ev_F \left( v_0(\mathbf{q}) f_3^{\pm}(\mathbf{q}) + v_1(\mathbf{q}) f_2^{\pm}(\mathbf{q}) + v_2(\mathbf{q}) f_1^{\pm}(\mathbf{q}) + v_3(\mathbf{q}) f_0^{\pm}(\mathbf{q}) \right) \pm$$
$$\pm 2ev_F \left( v_0(\mathbf{q}) f_2^{\pm}(\mathbf{q}) \frac{v_x^{\pm} \hbar}{k_B T}(1 - f_0^{\pm}(\mathbf{q})) + v_2(\mathbf{q}) f_0^{\pm}(\mathbf{q}) \frac{v_x^{\pm} \hbar}{k_B T}(1 - f_0^{\pm}(\mathbf{q})) \right) \mp$$
$$\mp 2ev_F \left( v_0(\mathbf{q}) f_1^{\pm}(\mathbf{q}) \frac{v_x^{\pm} \hbar}{k_B T} f_1^{\pm}(\mathbf{q}) + v_1(\mathbf{q}) f_0^{\pm}(\mathbf{q}) \frac{v_x^{\pm} \hbar}{k_B T} f_1^{\pm}(\mathbf{q}) + v_0(\mathbf{q}) f_0^{\pm}(\mathbf{q}) \frac{v_x^{\pm} \hbar}{k_B T} f_2^{\pm}(\mathbf{q}) \right)$$

The third harmonic generation coefficient can now be estimated as

$$\kappa_{13} = e^2 E_0^2 j_{x3}^{\pm} / (\hbar^2 \omega_0^2 j_{x1}^{\pm}). \tag{36}$$

Calculating the integrals for room temperature and $\mu_c = 0.1$ eV, we obtain a value for $E_0 = 10^6$ V/m and $\omega_0 = 10^{13}$ Hz, which is consistent with the experiment in [43,44].

## 5. Conclusions

As a conclusion, we note the following. The paper demonstrates the possibility of solving the problem of the nonlinear response of graphene in the classical approach using integration over the BZ while taking into account tensor nonzero effective masses of quasiparticles in the strong coupling model [1]. The approach can be extended to a linear dispersion model with integration over infinite Dirac cones. In this case, it is possible to analytically calculate all the integrals in terms of incomplete Fermi-Dirac integrals, but the result becomes cumbersome. The applied perturbation theory may well be used up to fields of the order of $10^6$ V/m. With stronger fields, you should solve the KBE and not use decompositions. Apparently, restrictions on the perturbation method arise for fields of the order of $10^8$ V/m. The resulting solution allows us to find a response in the form of $\sigma_{1xx}$ and $\sigma_{3xx}$ to the considered polarization. Similarly, considering orthogonal polarization, it is possible to find $\sigma_{1yy}$ and $\sigma_{3yy}$, i.e. the solution can be generalized to arbitrary polarization, taking into account the tensor character of the conductivity. Harmonic generation in a strong field is a non–linear and non-stationary process that requires solving the problem of heating up and changing $\mu_c(t)$ over time. It is also necessary to solve the non-stationary problem of graphene heating.

Graphene should actually be located on a substrate. Taking into account the substrate generally leads to a surface-volume integral equation. In the case of a sheet and a substrate infinite in two directions, the corresponding Green's function is known, and the solution can be reduced to a one-dimensional integral equation for the current and field in the substrate. In the case of a finite sheet, the task is complicated by the need to solve a combined integral equation with finding a current density satisfying the conditions at the edges. The dielectric substrate makes it possible to enhance harmonic generation. Since the harmonic amplitudes are small, the solution in the linear monochromatic approximation gives the reflection coefficient $R(\omega_0) = (1 - Y(\omega_0))/(1 + Y(\omega_0))$, where the normalized input conductivity on graphene is

$$Y(\omega_0) = \sqrt{\varepsilon}\, \frac{1 + i\sqrt{\varepsilon}\tan(\omega_0\sqrt{\varepsilon}d/c)}{\sqrt{\varepsilon} + i\tan(\omega_0\sqrt{\varepsilon}d/c)} + \eta_0 \sigma(\omega_0).$$

The linear conductivity is introduced here. The thickness of the substrate $d$ should be selected based on the condition of maximum absorbed power $P(\omega_0) = \mathrm{Re}(j^2(\omega_0)/\sigma(\omega_0))/2$. Due to the small conductivity of graphene, it is possible to approximate $k_0 d\sqrt{\varepsilon} \approx \pi/2$. In this case $Y(\omega) \approx \varepsilon$, and with a high dielectric constant we have $R(\omega) \approx -1$. This structure creates harmonics mainly in the reflected field. In thermodynamic equilibrium, graphene emits as much at all frequencies as it absorbs. The power absorbed by the main harmonic $P(\omega_0)$ is used to emit all odd harmonics, including the first one.

## APPENDIX

Using the dispersion (1), we have the decomposition

$$w(t,\mathbf{q}) = \sqrt{1 + 4\cos\left(\frac{q_y a}{2}\right)\cos\left(\sqrt{3}a\frac{q_x - q_0(t)}{2}\right) + 4\cos^2\left(\frac{q_y a}{2}\right)}$$

in the form

$$w(t,\mathbf{q}) = w(\mathbf{q}) + w_1(\mathbf{q})q_0(t) + w_2(\mathbf{q})q_0^2(t) + w_3(\mathbf{q})q_0^3(t) + \ldots,$$

$$w_1(\mathbf{q}) = \frac{\sqrt{3}a}{w(\mathbf{q})}\sin\left(\sqrt{3}a\frac{q_x}{2}\right)\cos\left(\frac{q_y a}{2}\right),$$

$$w_2(\mathbf{q}) = -\cos\left(\frac{q_y a}{2}\right)\frac{6a^2 \cos\left(\frac{q_y a}{2}\right)\sin^2\left(\sqrt{3}a\frac{q_x}{2}\right) + w^2(\mathbf{q})\cos\left(\sqrt{3}a\frac{q_x}{2}\right)}{8w^3(\mathbf{q})}, \qquad (A1)$$

$$w_3(\mathbf{q}) = \sqrt{3}a^3 \sin\left(\sqrt{3}a\frac{q_x}{2}\right)\cos\left(\frac{q_y a}{2}\right) \times$$

$$\times \frac{6a\cos\left(\frac{q_y a}{2}\right)\left[2\cos\left(\frac{q_y a}{2}\right)\sin^2\left(\sqrt{3}a\frac{q_x}{2}\right) + \cos\left(\sqrt{3}a\frac{q_x}{2}\right)w^2(\mathbf{q})\right] - w^4(\mathbf{q})}{8w^5(\mathbf{q})}.$$

The functions $w_1$ and $w_3$ are odd in $q_x$, and the function $w_2$ is even. With a small perturbation $q_0$ away from the Dirac points, $w$ is a positive value. In the vicinity of Dirac points $w_1(\mathbf{q_D}) = w_3(\mathbf{q_D}) = 0$, since the sine vanishes, whereas $w=0$ only at the points themselves, therefore the function $w$ is not negative (Fig. 3).

For the decomposition of the velocity $v_x^\pm(t,\mathbf{q})$, we have

$$v_x^\pm(t,\mathbf{q}) = \mp\left[v_0(\mathbf{q}) + v_1(\mathbf{q})q_0(t) + v_2(\mathbf{q})q_0^2(t) + v_3(\mathbf{q})q_0^3(t)\right],$$

$$v_0(\mathbf{q}) = v_F \frac{2\cos\left(a\frac{q_y}{2}\right)\sin\left(\sqrt{3}a\frac{q_x}{2}\right)}{w(\mathbf{q})},$$

$$v_1(\mathbf{q}) = v_F \frac{2\cos\left(a\frac{q_y}{2}\right)\left[s_1 w(\mathbf{q}) - w_1(\mathbf{q})\sin\left(\sqrt{3}a\frac{q_x}{2}\right)\right]}{w^2(\mathbf{q})}, \qquad (A2)$$

$$v_2(\mathbf{q}) = v_F \frac{2\cos\left(\frac{q_y a}{2}\right)\left[s_2 w(\mathbf{q}) - s_1 w_1(\mathbf{q}) - \sin\left(\frac{q_x\sqrt{3}a}{2}\right)\left(w_2(\mathbf{q}) - \frac{w_1^2(\mathbf{q})}{w(\mathbf{q})}\right)\right]}{w^2(\mathbf{q})},$$

$$v_3(\mathbf{q}) = v_F \frac{2\cos\left(\frac{q_y a}{2}\right)\left[s_3 w(\mathbf{q}) - s_2 w_1(\mathbf{q}) - s_1\left(w_2(\mathbf{q}) - w_1^2(\mathbf{q})/w(\mathbf{q})\right) - \sin\left(\frac{q_x\sqrt{3}a}{2}\right)w_3(\mathbf{q})\right]}{w^2(\mathbf{q})}.$$

In them $s_1 = -(\sqrt{3}a/2)\cos(\sqrt{3}aq_x/2)$, $s_2 = -(3a^2/8)\sin(\sqrt{3}aq_x/2)$, $s_3 = (\sqrt{9}a^3/16)\cos(\sqrt{3}aq_x/2)$ are the coefficients of the sine expansion:

$$\sin\left(\sqrt{3}a\frac{q_x - q_0(t)}{2}\right) = \sin\left(\sqrt{3}a\frac{q_x}{2}\right) + s_1 q_0(t) + s_2 q_0^2(t) + s_3 q_0^3(t).$$

The functions $v_0$ and $v_2$ are odd in $q_x$, and the functions $v_1$ and $v_3$ are even. All functions are even in $q_y$.

The decomposition of the exponent included in the DF $f_e^\pm(t,\mathbf{q}) = [1+\exp(u_\pm(t,\mathbf{q}))]^{-1}$ has the form

$$\exp(u_\pm(\mathbf{q})+\delta_\pm) = \exp(u_\pm(\mathbf{q}))[1+\delta_\pm+\delta_\pm^2/2+\delta_\pm^3/6],$$

$$u_\pm(t,\mathbf{q}) = \pm(\gamma_0 w(t,\mathbf{q})-\mu_c)/(k_B T) = u_\pm(\mathbf{q})+\delta_\pm(t,\mathbf{q}),$$

$$\delta_\pm(t,\mathbf{q}) = \pm\gamma_0(w_1(\mathbf{q})q_0(t)+w_2(\mathbf{q})q_0^2(t)+w_3(\mathbf{q})q_0^3(t))/(k_B T),$$

$$\exp(u_\pm+\delta_\pm) = \exp(u_\pm) \pm \frac{\gamma_0 w_1(\mathbf{q})}{k_B T}q_0(t) + \left(\frac{\gamma_0^2 w_1^2(\mathbf{q})}{2(k_B T)^2} \pm \frac{\gamma_0 w_2(\mathbf{q})}{k_B T}\right)q_0^2(t)$$
$$+\left(\frac{\gamma_0^2 2w_1(\mathbf{q})w_2(\mathbf{q})}{2(k_B T)^2} \pm \frac{\gamma_0 w_3(\mathbf{q})q_0^3(t)}{k_B T} + \frac{\gamma_0^3 w_1^2(\mathbf{q})}{6(k_B T)^3}\right)q_0^3(t).$$

Decomposing the fraction, we get the decomposition of DF

$$f_0^\pm(t,\mathbf{q}) = f_0^\pm(\mathbf{q}) + f_1^\pm(\mathbf{q})q_0 + f_2^\pm(\mathbf{q})q_0^2 + f_3^\pm(\mathbf{q})q_0^3,$$

$$f_0^\pm(\mathbf{q}) = \frac{1}{1+\exp(u_\pm(\mathbf{q}))},$$

$$f_1^\pm(\mathbf{q}) = -f_0^{\pm 2}(\mathbf{q})e_1^\pm(\mathbf{q}), \tag{A3}$$

$$f_2^\pm(\mathbf{q}) = -f_0^{\pm 2}(\mathbf{q})e_2^\pm + f_0^{\pm 3}(\mathbf{q})e_1^{\pm 2},$$

$$\left[f_{e3}^\pm(\mathbf{q}) = -f_0^{\pm 2}(\mathbf{q})e_3^\pm(\mathbf{q}) + 2f_0^{\pm 3}(\mathbf{q})e_1^\pm(\mathbf{q})e_2^\pm(\mathbf{q}) - f_0^{\pm 4}(\mathbf{q})e_1^{\pm 3}(\mathbf{q})\right],$$

the parity of which coincides with the parity of the indices, as well as the coefficients of the exponential expansion:

$$e_1^\pm(\mathbf{q}) = \pm\frac{\gamma_0 w_1(\mathbf{q})}{k_B T},$$

$$e_2^\pm(\mathbf{q}) = \left(\frac{\gamma_0^2 w_1^2(\mathbf{q})}{2(k_B T)^2} \pm \frac{\gamma_0 w_2(\mathbf{q})}{k_B T}\right), \tag{A4}$$

$$e_3^\pm(\mathbf{q}) = \left(\frac{\gamma_0^2 2w_1(\mathbf{q})w_2(\mathbf{q})}{2(k_B T)^2} \pm \frac{\gamma_0 w_3(\mathbf{q})q_0^3(t)}{k_B T} + \frac{\gamma_0^3 w_1^2(\mathbf{q})}{6(k_B T)^3}\right).$$